\documentclass[journal]{IEEEtran}

\usepackage{cite}
\usepackage{amsmath,amssymb,amsfonts,commath}
\DeclareMathOperator{\sgn}{sgn}
\usepackage{algorithmic}
\usepackage{graphicx}
\usepackage{subcaption}
\usepackage{textcomp}
\usepackage{xcolor}
\usepackage{array}
\usepackage{url}
\usepackage{stfloats}

\begin{document}
\title{A Novel SEPIC-\'Cuk Based High Gain Solar Micro-Inverter for Grid Integration}
\author{Arup~Ratan~Paul,
        Arghyadip~Bhattacharya,
        and~Kishore~Chatterjee,~\IEEEmembership{Member,~IEEE}% 
%\thanks{}% <-this % stops a space
}
%\IEEEspecialpapernotice{(Invited Paper)}

%\markboth{IEEE TRANSACTIONS ON INDUSTRY APPLICATIONS}%
%{IEEE TRANSACTIONS ON INDUSTRY APPLICATIONS}

\maketitle

\begin{abstract}
Solar micro-inverters are becoming increasingly popular as they are modular, and they posses the capability of extracting maximum available power from the individual photovoltaic (PV) modules of a solar  array. For realizing micro-inverters single stage transformer-less topologies are preferred as they offer better power evacuation efficacy. A SEPIC-\'{C}uk based transformer-less micro-inverter, having only one high frequency switch and four line frequency switches, is proposed in this paper. The proposed converter can be employed to interface a 35~V PV module to a 220~V single phase ac grid. As a very high gain is required to be achieved for the converter, it is made to operate in discontinuous conduction mode (DCM) for all possible operating conditions. Since the ground of the each PV modules is connected to the ground of the utility, there is no possibility of leakage current flow between the module and the utility. Detailed simulation studies are carried out to ascertain the efficacy of the proposed micro-inverter. A laboratory prototype of the inverter is fabricated, and detailed experimental studies are carried out to confirm the viability of the proposed scheme.
\end{abstract}

% Note that keywords are not normally used for peerreview papers.

% For peer review papers, you can put extra information on the cover
% page as needed:
% \ifCLASSOPTIONpeerreview
% \begin{center} \bfseries EDICS Category: 3-BBND \end{center}
% \fi
%
% For peerreview papers, this IEEEtran command inserts a page break and
% creates the second title. It will be ignored for other modes.
\IEEEpeerreviewmaketitle

\section{Introduction}

\IEEEPARstart{T}{hough} the fossil fuel based power generation remains to be the dominant source of electrical energy, harvesting of energy from the renewable sources (solar, wind etc.) are strongly being encouraged as they are more environment friendly, and inexhaustible in nature. In India, as the solar energy is abundant all over the country throughout the year, it is becoming a viable alternative to the conventional methods of generation of electrical energy. The Ministry of New \& Renewable Energy, Govt. of India, has set a target of 100 GW grid connected solar power generation by the year 2021-22 under National Solar Mission, out of which 40 GW should be from the rooftop installations~\cite{MNRE}. The solar modules are generally interfaced to the utility by employing the following configurations, (i) central inverter, (ii) string inverter and (iii) micro-inverter. The major advantages of the micro-inverter over the central and string inverter are as follows~\cite{bidram2012control},
\begin{itemize}
	\item capability of individual maximum power point tracking (MPPT) of the PV modules during non-uniform shading,
	\item the aspect of modularity bestows it with the flexibility in future expansion of the plant whenever required,
	\item the existence of plug and play feature imparts flexibility in operation and coordination of the system.
\end{itemize}

Typically, the power rating of a micro-inverter is 250-350~W, and hence their efficiency remains to be low compared to that of the schemes based on central and string inverters. In order to improve the efficiency of micro-inverters, lower number of power conversion stages are incorporated~\cite{meneses2013review}. In order to accomplish this feature, transformer-less single stage topologies for micro-inverter are generally preferred~\cite{8648358}. However, in case of transformer-less topologies, there exists a path for the leakage current to flow through the parasitic capacitance of the PV module to the grid~\cite{gubia2007ground,debnath2015solar}. The flow of leakage current deteriorates the life of the modules. Further, it posses hazard to the working personnel as the potential of the mountings of the PV module may become high with respect to the mother earth. Hence, the magnitude of the leakage current needs to be limited within the standard limit as stipulated by various regulatory authorities~\cite{Wli2015Leakage}.

Generally the MPP voltage of a PV module is 35~V. However it has to be interfaced to the single phase 110/230~V grid. Hence, the micro-inverter needs to have a very high gain, which is one of the main challenges to be addressed.

The requirements pertaining to topological configuration, and its design can be summarized as follows,
\begin{itemize}
	\item the magnitude of the leakage current must be within the limits as specified by various standards,
	\item the inverter system should have a very high voltage gain to interface a PV module having a $V_{mpp}$ of 35~V to a single phase 110/230~V ac grid,
	\item the micro-inverter needs to made as compact as possible.
%	\item High efficiency as much as possible. 
\end{itemize}

There are several single stage non-isolated topologies reported in the literature \cite{8648358}. The issue related to the leakage current has been addressed to by shorting the module ground with that of the ac grid~\cite{patel2009,jamatia2016single,gautam2017design,7558143,jamatia2018cuk,8013707,rajeev2018analysis,sarikhani2020threeswitch}. A buck-boost based micro-inverter topology with only one high frequency switch is proposed in \cite{patel2009}. In~\cite{jamatia2016single}, a buck-boost based and in~\cite{gautam2017design}, a \'{C}uk converter based single stage micro-inverter topologies are reported, wherein the PV module and the ac grid shares the common ground. In~\cite{jamatia2016single}, the MPP voltage is chosen to be 55~V, and the ac grid voltage is considered to be 110~V. Though the converter is operated in continuous conduction mode (CCM), the source current is discontinuous thereby increasing the filtering requirement at the input side. Moreover, the switches are operated in such a way that, the decoupling of second order harmonic component is achieved at the terminals of the PV module. In~\cite{gautam2017design}, \'Cuk derived topology is reported wherein, 5 high frequency (HF) switches are employed, and the converter is operated in CCM. A similar power decoupling strategy, as reported in \cite{jamatia2016single}, is incorporated in the topology presented in\cite{gautam2017design}, and reported in~\cite{jamatia2018cuk}. The designed switching frequency in all the aforesaid configurations are 50~kHz. In \cite{7558143} and \cite{8013707}, a four switch topology with different switching configurations have been proposed. However, all the switches of these converters \cite{jamatia2016single,gautam2017design,7558143,8013707} operate at high frequency. All these converters are designed to be operated in CCM. This facilitates the inverters to get interfaced to a single phase voltage level up to 110~V. A modification to the circuit reported in \cite{gautam2017design} has been accomplished by adding another HF switch for controlling the reactive power fed to the grid, and is presented in \cite{rajeev2018analysis}. However, the designed switching frequency is 10~kHz, and hence the size of passive elements are significantly large.  A three switch micro-inverter topology with common ground is reported in \cite{sarikhani2020threeswitch} wherein the voltage gain of the converter can be made both positive and negative, and therefore it does not require the service of an unfolding circuit but then it involves three inductors in the main converter.

In India, the ac distribution voltage level is 220~V. However, all the aforementioned converters are designed to get interfaced to single phase 110~V system, and hence they are not suitable for Indian conditions. However in \cite{8980872} a micro-inverter topology is presented, which can be interfaced to a single phase 220~V grid. However, it requires six switches, out of which, two are required to be high frequency switches. In view of this an effort has been made in this paper to design a transformer-less micro-inverter suitable for getting interfaced to a single phase 220~V grid while utilizing only one high frequency switch and four line frequency switches in order to reduce the switch count and reliability of the system.

In order to accomplish these features a SEPIC-\'Cuk based micro-inverter topology is proposed in \cite{9034790} by the authors of the current paper. The converter is operated in SEPIC mode during positive half cycle and in \'Cuk mode during negative half cycle. By merging SEPIC and \'Cuk configurations into one topology, the number of HF switches has been reduced to one. The designed switching frequency of the system is 100~kHz, hence the size of passive elements have been reduced considerably. The presence of the common ground between the module and the grid ensures negligible leakage current flow. However, the circuit was explored for getting interfaced to single phase 110~V ac grid. In this paper, an attempt has been made to get interfaced to a single phase 220~V ac, which is suitable for Indian distribution network, utilizing the converter proposed in \cite{9034790} from a PV module having $V_{mpp}$ of 35~V. Detailed simulation studies are carried out on MATLAB/Simulink platform to ascertain the effectiveness of the proposed converter. A semi-engineered laboratory prototype of the inverter has been fabricated, and detailed experimental studies are carried out to confirm the viability of the proposed scheme.

\section{Principle of Operation of the Proposed Topology}
\begin{figure}
	\includegraphics[width=\linewidth]{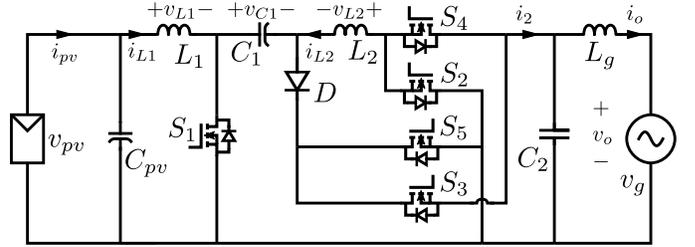}
	\caption{Schematic diagram of \'{C}uk-SEPIC based micro-inverter with dc side filter}
	\label{fig:cuksepic}
\end{figure}

The schematic diagram of the proposed SEPIC-\'Cuk based micro-inverter is shown in Fig.~\ref{fig:cuksepic}. Four line frequency switches are utilized to derive the combine effect of SEPIC and \'Cuk configuration. In \cite{9034790} the IGBTs are used to realize the line frequency switches. In this paper MOSFETs are used instead of IGBTs to reduce the losses. The diode, $D$ is employed to block the reverse current flow through the four MOSFETs. As sinusoidal current is required to be injected to the grid, the high frequency switch is applied with switching pulses which are obtained by comparing the rectified sine wave with a triangular carrier wave. The instantaneous duty ratio of $S_1$ is determined by multiplying a rectified sine wave with the peak duty ratio, $D_{peak}$. The switch is operated with $D_{peak}$ when the instantaneous grid voltage is at its peak. The current through $L_2$ is a rectified sine wave having high frequency switching harmonics superimposed on it. In order to transform this rectified current into a sinusoidal waveform the four switches, $S_2, S_3, S_4$, and $S_5$ are appropriately switched. The output capacitor, $C_2$ filters out the high frequency harmonic components form the output current.

The inductors of the converter can be chosen in such a way so that the converter can be operated either in CCM or in DCM. In CCM the voltage gain of both SEPIC and \'Cuk converter is $\frac{d}{1-d}$, wherein $d$ is the duty ratio of the converter. However, for interfacing the 35~V module to a 220~V ac grid whose peak voltage is 311~V, the requirement of the duty ratio at the peak of the ac grid voltage is required to be maintained at 0.9. Operating the inverter at this high duty ratio, would reduce the operating efficiency significantly. In order to overcome this limitation, the proposed topology is operated in DCM to obtain high voltage gain while maintaining reduced duty ratio compared to that of CCM operation. In case of a SEPIC and \'Cuk converter, current through both $L_1$ and $L_2$ does not become zero either in CCM or DCM mode of operation\cite{shi2013,5136983}. %In DCM, the voltages across these two inductors become zero for the interval which starts when the sum of $i_{L1}$ and $i_{L2}$. The value of $i_{L1}$ and $i_{L2}$ may not be zero, but remains constant for this interval.
Further, the DCM operation ensures that turn on switching loss of $S_1$ is negligible. But this is achieved at the cost of increased peak current that flows through the input inductor.% As the inductor ($L_1$) current is discontinuous, input current contains switching frequency harmonics, which can be eliminated by designing proper input filter (Fig. \ref{fig:cuksepic}).

\section{Modes of Operation and Analysis}
In order to simplify the analysis of the inverter, a dc source, $V_{dc}$ is used instead of a PV module. The convention of voltage polarity and current direction is shown in Fig.~\ref{fig:cuksepic}. Since the switching frequency is very high compared to that of the line frequency, it is assumed that, the output voltage and current requirement is almost constant throughout the switching time period ($T_s$). The ripple in the voltages in the capacitors, $C_1$ and $C_2$ can be considered to be negligible. The average value of $v_{C1}$ and $v_o$ over a switching time period are assumed to be $V_{C1}$ and $V_o$ respectively. The current and voltage waveforms over a switching cycle is shown in Fig.~\ref{fig:wavefrorm}. The switching time period ($T_s$) is divided into three time intervals as follows, (i) $T_1(=DT_s)$, (ii) $T_2$ and (iii) $T_3(=D_0 T_s)$ respectively. As $T_1+T_2+T_3=T_s$, therefore $T_2=(1-D-D_0)T_s$.

\begin{figure*}[!t]
	\centering
	\begin{subfigure}{0.32\linewidth}
		\includegraphics[width=\linewidth]{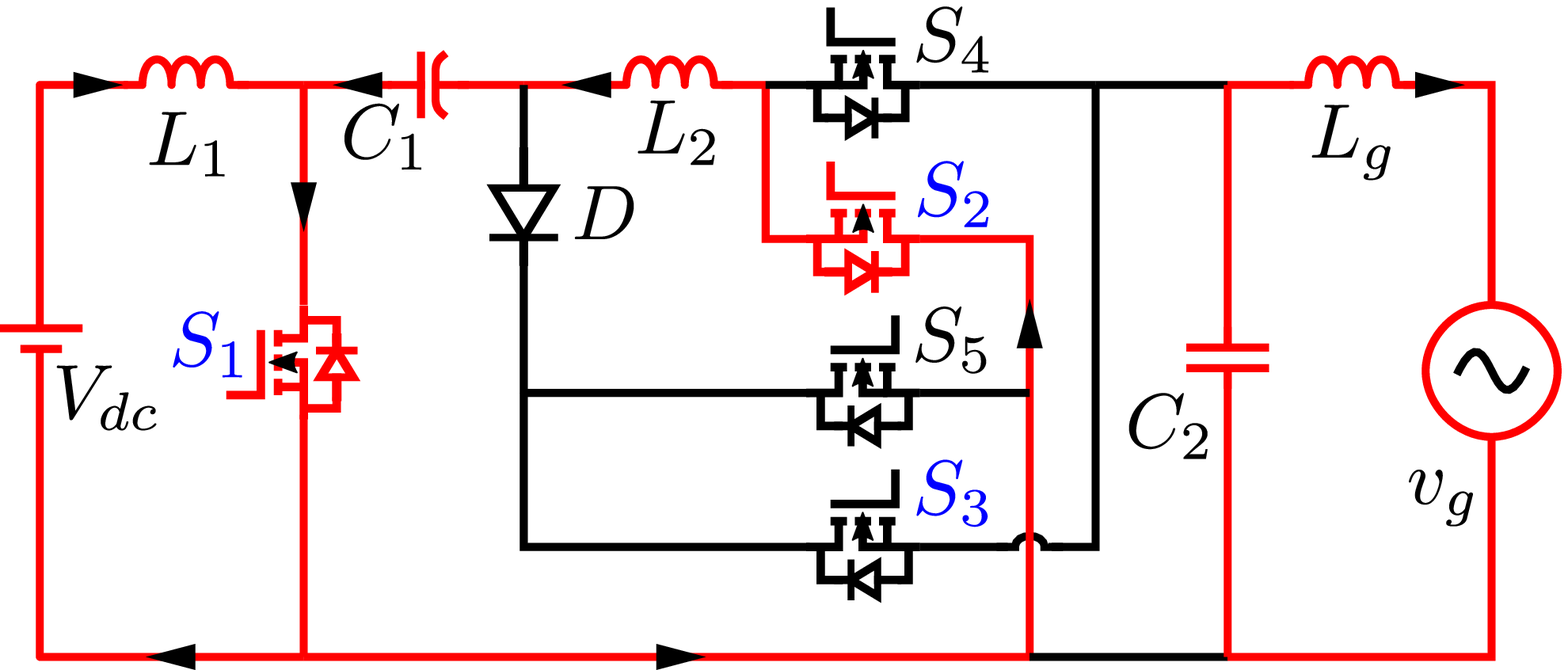}
		\subcaption{}
	\end{subfigure}
\hfil
	\begin{subfigure}{0.32\linewidth}
		\includegraphics[width=\linewidth]{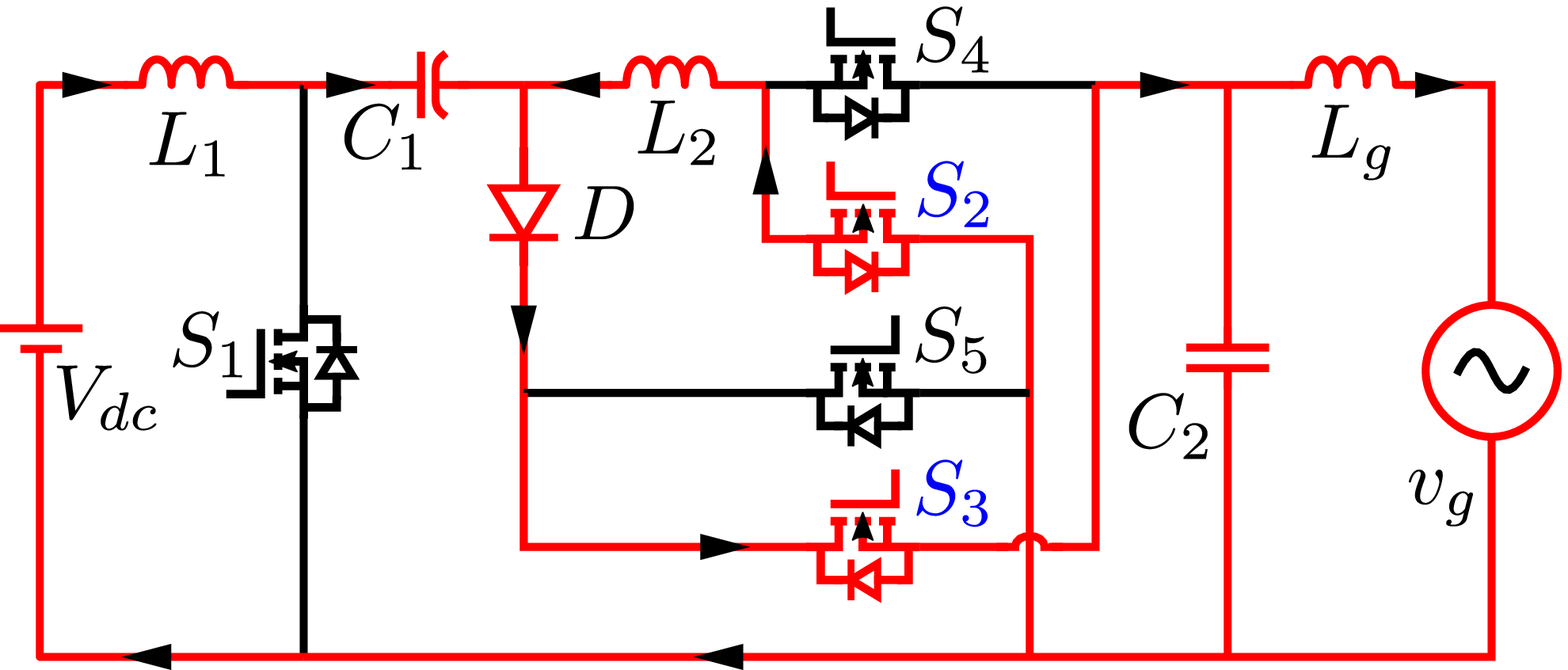}
		\subcaption{}
	\end{subfigure}
\hfil
	\begin{subfigure}{0.32\linewidth}
		\includegraphics[width=\linewidth]{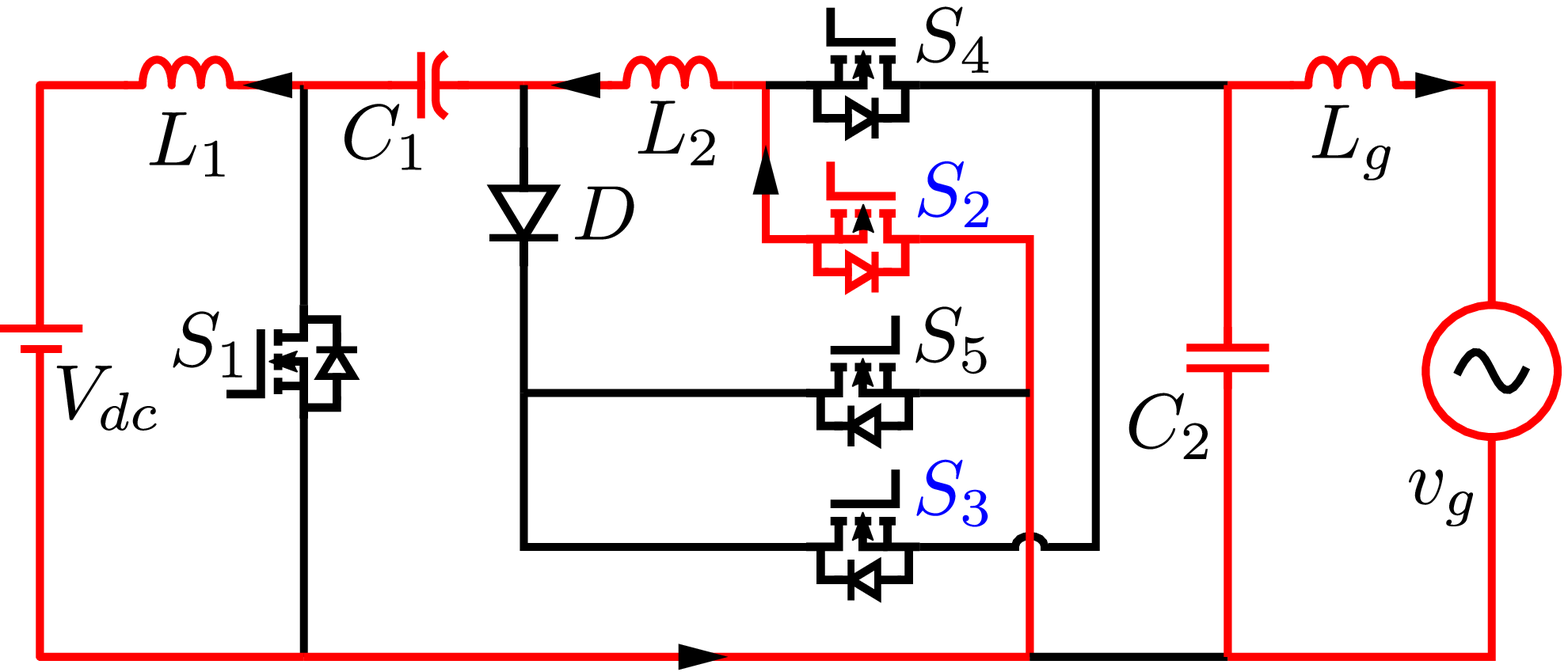}
		\subcaption{}
	\end{subfigure}
	\caption{Modes of operation for positive half cycle: (a) Mode-I, (b) Mode-II, and (c) Mode-III}
	\label{fig:cuksepic_phc}
	\vfil
	\begin{subfigure}{0.32\linewidth}
		\centering
		\includegraphics[width=\linewidth]{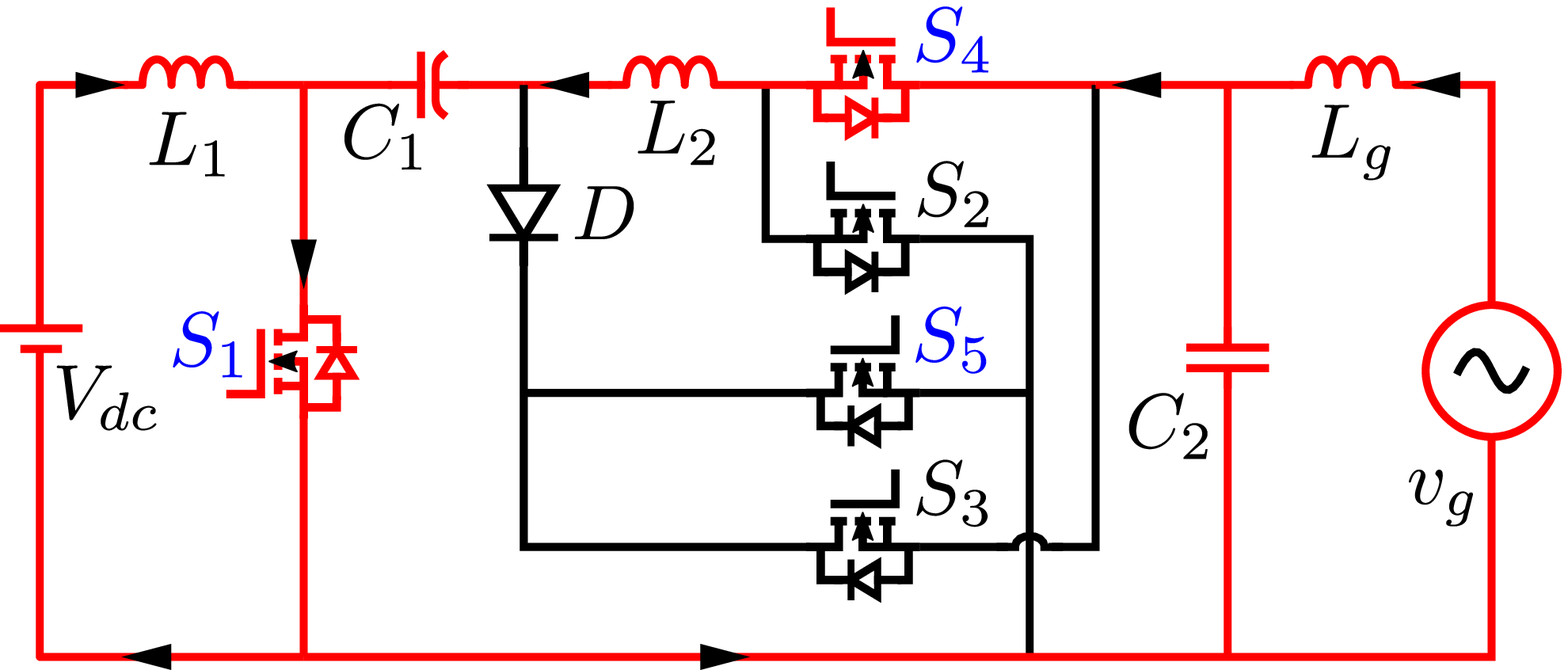}
		\subcaption{}
	\end{subfigure}
\hfil
	\begin{subfigure}{0.32\linewidth}
		\centering
		\includegraphics[width=\linewidth]{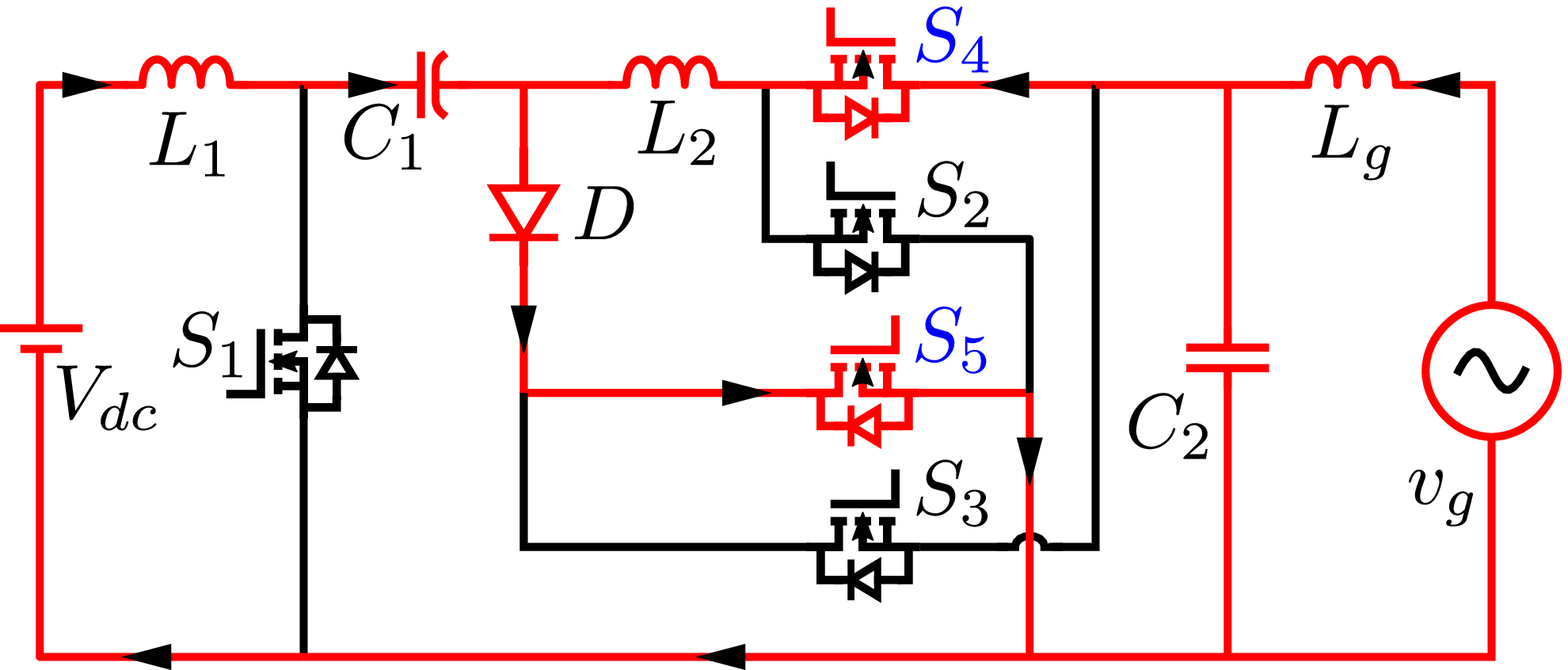}
		\subcaption{}
	\end{subfigure}
\hfil
	\begin{subfigure}{0.32\linewidth}
		\centering
		\includegraphics[width=\linewidth]{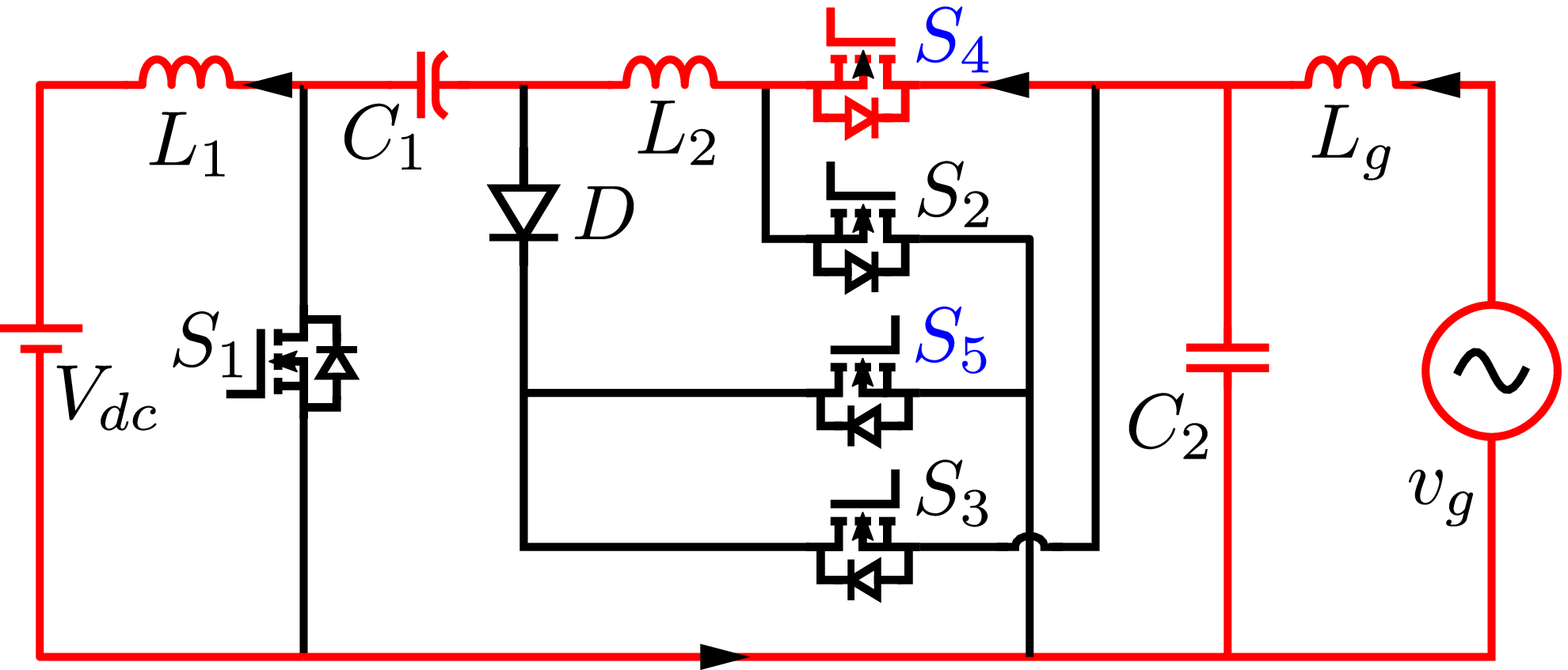}
		\subcaption{}
	\end{subfigure}
	\caption{Modes of operation for negative half cycle: (a) Mode-I, (b) Mode-II, and (c) Mode-III}
	\label{fig:cuksepic_nhc}
\end{figure*}

\subsection{Positive half cycle (SEPIC mode)}

Switch $S_2$ and $S_3$ both are turned on, while $S_4$ and $S_5$ are kept off. At steady state, as the switching time average voltages across $L_1$ and $L_2$ are zero,
\begin{align}
	V_{C1} &= V_{dc}
	\label{eqn:7}
\end{align}

\subsubsection{Mode I  ($0\leq t < T_1$)}
$S_1$ is turned on and the current through $L_1$ increases as it flows through the path, $V_{dc}-L_1-S_1-V_{dc}$ as shown in Fig.~\ref{fig:cuksepic_phc}(a). The intermediate capacitor, $C_1$ gets discharged while the current flowing through $L_2$ increases as it flows through the path, $C_1-S_1-S_2-L_2-C_1$. The capacitor, $C_2$ supplies power to the grid in this interval. The diode, $D$ blocks the current through the switch $S_3$ even though it is receiving the gating pulse. Hence,
	\begin{align}
		V_{dc} &= L_1\frac{di_{L1}}{dt}=L_1\frac{I_{L1p}-I_{L1v}}{DT_s}=L_1\frac{\Delta I_{L1}}{DT_s}
		\label{eqn:1} \\
		V_{C1} &= L_2\frac{di_{L2}}{dt}=L_2\frac{I_{L2p}-I_{L2v}}{DT_s}=L_2\frac{\Delta I_{L2}}{DT_s} 
		\label{eqn:2} \\
		i_2&=0 
		\label{eqn:3}
	\end{align}%
where, $\Delta I_{L1}=(I_{L1p}-I_{L1v})$ and $\Delta I_{L2}=(I_{L2p}-I_{L2v})$.
	\begin{figure}
		\centering
		\begin{subfigure}{0.42\linewidth}
			\centering
			\includegraphics[width=\linewidth]{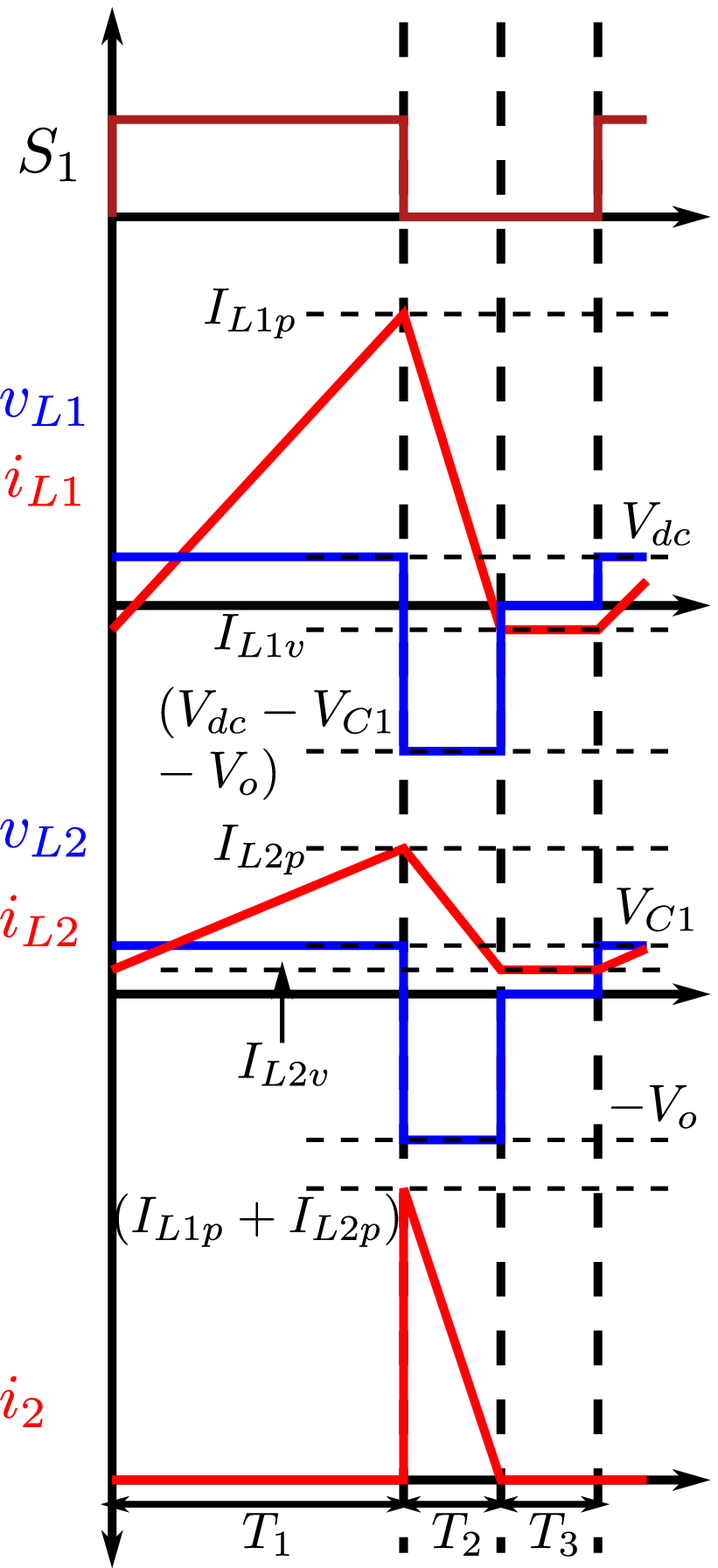}
			\subcaption{}
		\end{subfigure}
		\hfil
		\begin{subfigure}{0.42\linewidth}
			\centering
			\includegraphics[width=\linewidth]{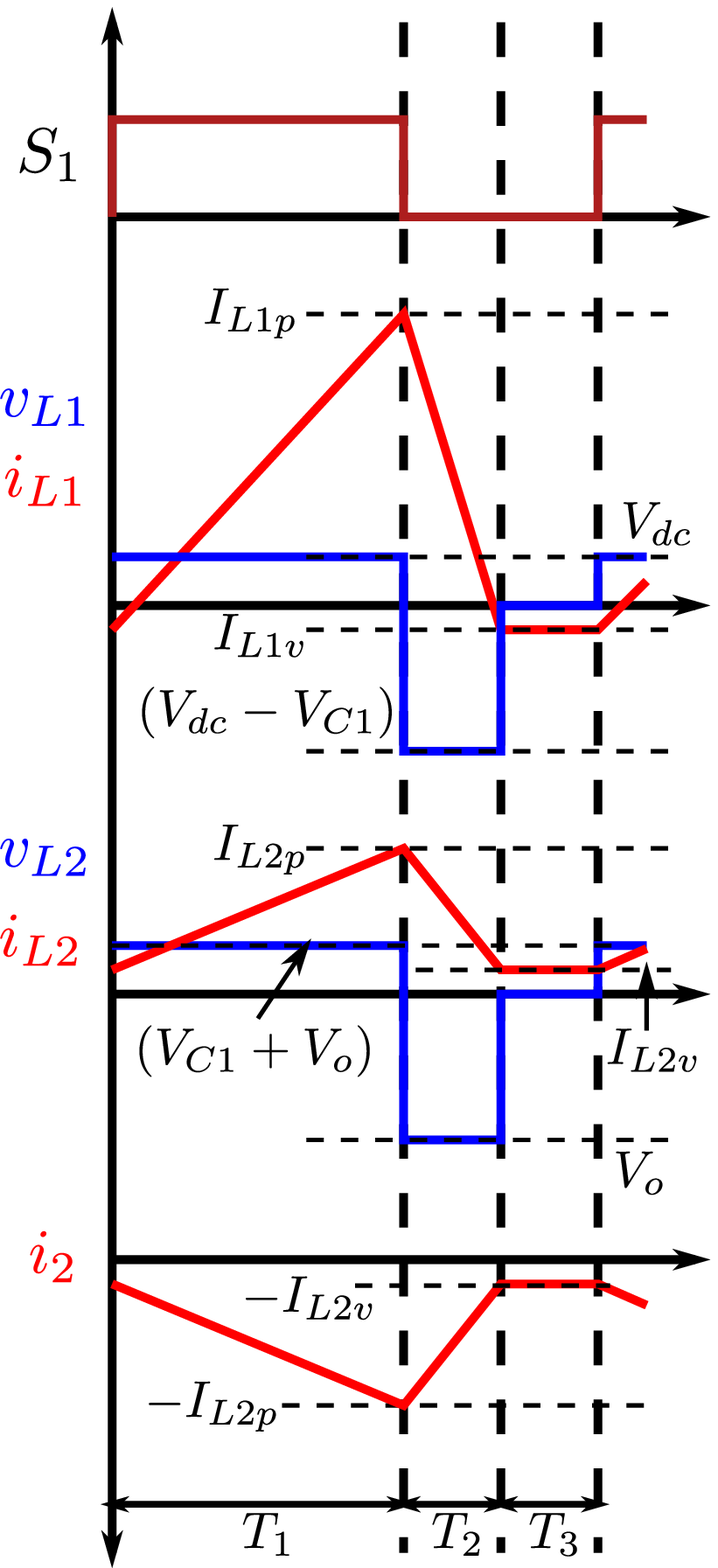}
			\subcaption{}
		\end{subfigure}%\vspace{3mm}
		\caption{Waveform of different converter parameters over a switching time period: (a) for positive half cycle and (b) for negative half cycle}
		\label{fig:wavefrorm}
	\end{figure}
\subsubsection{Mode II ($T_1\leq t < (T_1+T_2)$)}
$S_1$ is turned off. The current flows through the path, $V_{dc}-L_1-C_1-D-S_3-Load-V_{dc}$ as shown in Fig.~\ref{fig:cuksepic_phc}(b) thereby charging $C_1$. The energy stored in $L2$ is transferred to the grid as well as to $C_2$ through the path, $L_2-D-S_3-Load-S_2-L_2$. This mode ends when $i_{L1}=-i_{L2}$. At the end of this mode, let the magnitude of the currents through $L_1$ and $L_2$ be $I_{L1m}$ and $I_{L2m}$ respectively. Therefore,
	\begin{align}
		V_{dc}-V_{C1}-V_o&=L_1\frac{di_{L1}}{dt}=L_1\frac{I_{L1m}-I_{L1p}}{(1-D-D_0)T_s}
		\label{eqn:4}\\
		-V_o&=L_2\frac{di_{L2}}{dt}=L_2\frac{I_{L2m}-I_{L2p}}{(1-D-D_0)T_s}
		\label{eqn:5}\\
		i_2&=i_{L1}+i_{L2}.
		\label{eqn:6}
	\end{align}
\subsubsection{Mode III ($(T_1+T_2)\leq t < T_s$)}
This mode starts when the current through the diode, $D$ becomes zero. The diode turns off and blocks the current flowing through $S_3$. The current path in this mode is $V_{dc}-S_2-L_2-C_1-L_1-V_{dc}$ as shown in Fig.~\ref{fig:cuksepic_phc}(c). Since $V_{C1}=V_{dc}$, the currents in the inductors will remain unchanged in this mode. The capacitor $C_2$ supplies power to the grid in this interval. Hence,
	\begin{align}
		I_{L1m}&=I_{L1v}=-I_{L2m}=-I_{L2v}.
		\label{eqn:mv1}
	\end{align}
	%	\begin{equation*}
	%		V_{dc}-V_{C1}=(L_1+L_2)\frac{di_{L1}}{dt}
	%	\end{equation*}
	%	\begin{equation}
	%		\implies V_{dc}-V_{C1}=(L_1+L_2)\frac{I_{L1v}-I_{L1m}}{(1-D-D_0)T_s}.
	%	\label{eqn:7b}
	%	\end{equation}	
	%	It is to be noted that, in this interval, $i_{L1}+i_{L2}=0$.

Solving (\ref{eqn:1}), (\ref{eqn:2}), (\ref{eqn:4}), (\ref{eqn:5}) and (\ref{eqn:mv1}), the expression for the voltage gain can be obtained as,
%	\begin{equation}
%		V_{C1}=V_{dc}
%	\end{equation}
%	and
%	\begin{equation}
%		I_{L1m}=I_{L2v}, \quad I_{L2m}=I_{L2v}
%	\end{equation} which means the inductor currents are constant in the mode-III.

\begin{align}
	\frac{V_o}{V_{dc}}=\frac{D}{\del{1-D-D_0}}.
	\label{eqn:pgain}
\end{align}

\subsection{Negative half cycle (\'{C}uk Mode):}
Switch $S_4$ and $S_5$ both are turned on, while $S_2$ and $S_3$ are kept off. At steady state, as the switching time average voltages across $L_1$ and $L_2$ are zero,
\begin{align}
	V_{C1}&=V_{dc}-V_o.
	\label{eqn:14}
\end{align}
\subsubsection{Mode I ($0\leq t < T_1$)}
$S_1$ is turned on and the current through $L_1$ increases as it flows through the path, $V_{dc}-L_1-S_1-V_{dc}$ as shown in Fig.~\ref{fig:cuksepic_nhc}(a). The intermediate capacitor, $C_1$ gets discharged while the current flowing through $L_2$ increases as it flows through the path, $C_1-S_1-Load-S_4-L_2-C_1$. The diode, $D$ blocks the current through the switch $S_5$ even though it is receiving the gating pulse. Hence,
	\begin{align}
		V_{dc}&=L_1\frac{di_{L1}}{dt}=L_1\frac{I_{L1p}-I_{L1v}}{DT_s}
		\label{eqn:9}\\
		V_{C1}+V_o&=L_2\frac{di_{L2}}{dt}=L_2\frac{I_{L2p}-I_{L2v}}{DT_s}.
		\label{eqn:10}
	\end{align}
\subsubsection{Mode II ($T_1\leq t < (T_1+T_2)$)}
$S_1$ is turned OFF. The current flows through the path, $V_{dc}-L_1-C_1-D-S_5-V_{dc}$ as shown in Fig.~\ref{fig:cuksepic_nhc}(b) thereby charging $C_1$. The energy stored in $L2$ is transferred to the load as well as to $C_2$ through the path, $L_2-D-S_5-Load-S_4-L_2$. This mode ends when $i_{L1}=-i_{L2}$. At the end of this mode, let the magnitude of the currents through $L_1$ and $L_2$ be $I_{L1m}$ and $I_{L2m}$ respectively. Therefore,
	\begin{align}
		V_{dc}-V_{C1}=L_1\frac{di_{L1}}{dt}=L_1\frac{I_{L1m}-I_{L1p}}{(1-D-D_0)T_s}
		\label{eqn:11}\\
		V_o=L_2\frac{di_{L2}}{dt}=L_2\frac{I_{L2m}-I_{L2p}}{(1-D-D_0)T_s}.
		\label{eqn:12}
	\end{align}
	%	From equation \ref{eqn:9}, \ref{eqn:10}, \ref{eqn:11}, and \ref{eqn:12},
	%	\begin{equation}
	%	\frac{D}{1-D-D_0}=\frac{V_{C1}-V_{dc}}{V_{dc}}=\frac{-V_o}{V_{C1}+V_o}.
	%	\label{eqn:2a}
	%	\end{equation}
	%	\begin{equation}
	%	\implies V_{C1}=V_{dc}-V_o
	%	\label{eqn:13}
	%	\end{equation}
	%	since $V_{C1}\neq0$.
	%	From equation \ref{eqn:2a} and \ref{eqn:13},
	%	\begin{equation}
	%	\frac{V_o}{V_{dc}}=-\frac{D}{1-D-D_0}.
	%	\label{eqn:ngain}
	%	\end{equation}
\subsubsection{Mode III ($(T_1+T_2)\leq t < T_s$)}
This mode starts when the current through the diode, $D$ becomes zero. The diode turns off and blocks the current flowing through $S_5$. The current path in this mode is $V_{dc}-Load-S_4-L_2-C_1-L_1-V_{dc}$ as shown in Fig.~\ref{fig:cuksepic_nhc}(c). Since $V_{C1}=V_{dc}-V_o$, the currents in the inductors will remain unchanged in this mode. Hence,
	\begin{align}
		I_{L1m}&=I_{L1v}=-I_{L2m}=-I_{L2v}.
		\label{eqn:mv2}
	\end{align}
	%	\begin{equation*}
	%		V_{dc}-V_{C1}-V_o=(L_1+L_2)\frac{di_{L1}}{dt}
	%	\end{equation*}
	%	\begin{equation}
	%		\implies V_{dc}-V_{C1}-V_o=(L_1+L_2)\frac{I_{L1v}-I_{L1m}}{(1-D-D_0)T_s}.
	%		\label{eqn:12a}
	%	\end{equation}	
	%	It is to be noted that, in this interval, $i_{L1}+i_{L2}=0$.
	%	\begin{equation}
	%		\implies I_{L1m}=-I_{L2m}, \quad I_{L1v}=-I_{L2v}.
	%	\end{equation}
Solving \eqref{eqn:9}--\eqref{eqn:mv2},  the expression for the voltage gain can be obtained as,
\begin{align}
	\frac{V_o}{V_{dc}}&=-\frac{D}{\del{1-D-D_0}}.
	\label{eqn:ngain}
\end{align}
\vspace{-6mm}
\subsection{Combined operation}
From (\ref{eqn:pgain}) and (\ref{eqn:ngain}) the common expression for the voltage gain can be written as,
\begin{align}
	\frac{V_o}{V_{dc}}&=\sgn(V_o)\frac{D}{\del{1-D-D_0}}.
	\label{eqn:gain}
\end{align}
From \eqref{eqn:7} and \eqref{eqn:14}, it can be inferred that, the expression for $V_{C1}$ is different in the positive and in the negative half cycle. By substituting the expression of $V_{C1}$ from \eqref{eqn:7} and \eqref{eqn:14} to \eqref{eqn:1}, \eqref{eqn:2}, \eqref{eqn:4}, \eqref{eqn:5}, \eqref{eqn:9}, \eqref{eqn:10}, \eqref{eqn:11} and \eqref{eqn:12}, the model of the system can be simplified as follows:
\begin{align}
	L_1\frac{\Delta I_{L1}}{DT_s}&=L_2\frac{\Delta I_{L2}}{DT_s}=V_{dc}
	\label{eqn:15}\\
	L_1\frac{\Delta I_{L1}}{(1-D-D_0)T_s}&=L_2\frac{\Delta I_{L2}}{(1-D-D_0)T_s}=|V_o|.
	\label{eqn:16}
\end{align}

From Fig.~\ref{fig:wavefrorm}, in both the half cycles the expression of switching time average of the dc side current is obtained as,
\begin{align}
	I_{dc}&=I_{L1v}+\frac{\Delta I_{L1}}{2}(1-D_0).
	\label{eqn:17}
\end{align}

Further, from Fig.~\ref{fig:wavefrorm} it can be observed that in the positive half cycle the wave shape of $i_2$ in a switching time period is different from that in the negative half cycle. Let the switching time average of $i_2$ is $I_2$, and it can be approximated to be equal to the grid current, $i_o$. It can be noted that, this waveform is exactly negative to that of $i_{L2}$ in \'Cuk mode of operation. Hence, in this mode, $I_2$ is equal to the switching time average of $i_{L2}$ but with a negative sign. Therefore,
\begin{align}
	|I_2|_{Cuk}&=I_{L2v}+\frac{\Delta I_{L2}}{2}\del{1-D_0}.
	\label{eqn:18}
\end{align}

For SEPIC mode, $I_2$ can be expressed as,
\begin{align}
	|I_2|_{SEPIC}&= \frac{\del{1-D-D_0}}{2}(I_{L1p}+I_{L2p}) \nonumber \\
		&=\frac{\del{1-D-D_0}}{2}(\Delta I_{L1}+\Delta I_{L2}).
	\label{eqn:19}
\end{align}

Considering the system to be loss-less,
\begin{align}
	\frac{I_{dc}}{|I_2|_{SEPIC}}=\frac{V_o}{V_{dc}}=\frac{D}{\del{1-D-D_0}}.
\end{align}

Using (\ref{eqn:15}), (\ref{eqn:17}) and (\ref{eqn:19}),
\begin{align} 		\frac{(1-D-D_0)\frac{DT_sV_{dc}}{L_{eq}}}{2I_{L1v}+(1-D_0)\frac{DT_sV_{dc}}{L1}}=\frac{\del{1-D-D_0}}{D}
	\label{eqn:20}
\end{align}
where $L_{eq}=L_1||L_2$. Simplifying \eqref{eqn:20},
\begin{align}
	I_{L1v}=\frac{DT_sV_{dc}}{2}\sbr{\frac{D}{L_2}-\frac{\del{1-D-D_0}}{L_1}}. \label{eqn:21}
\end{align}

If, $I_{L1v}$ has to be positive,
\begin{align}
	\frac{L_2}{L_1}<\frac{D}{1-D-D_0}=\frac{V_o}{V_{dc}}
	\label{eqn:L1L2}
\end{align} needs to be satisfied. However in that case when $V_o$ becomes zero at the zero crossing instants as per \eqref{eqn:L1L2}, $L_2$ needs to assume a negative value which is not possible. Hence the value of $L_2$ is chosen so that, $I_{L1v}$ remains negative for all possible operating conditions.

In SEPIC mode of operation, the average current through $L_2$ is given by,
\begin{align}
	I_{L2}&=I_{L2v}+\frac{\Delta I_{L2}}{2}(1-D_0) \nonumber\\
	&=\frac{DT_sV_{dc}}{2}\sbr{\frac{1-D-D_0}{L_1}-\frac{D}{L_2}}+\frac{DT_sV_{dc}}{2L_2}(1-D_0) \nonumber\\
	&=(1-D-D_0)\frac{DT_sV_{dc}}{2L_{eq}} \nonumber\\
	&=\frac{\del{1-D-D_0}}{2}(\Delta I_{L1}+\Delta I_{L2}). \label{eqn:22}
\end{align}

From (\ref{eqn:19}) and (\ref{eqn:22}) it can be inferred that, $I_2$ is equal to $I_{L2}$ in SEPIC mode of operation as well. Hence in combined operation, $I_{L_2}=|I_2|$, which means that, $i_{L2}$ can be manipulated to control $i_o$. Since the absolute value of the output current expression is same in both the half cycles it can be inferred that (\ref{eqn:21}) and (\ref{eqn:L1L2}) are also valid for \'Cuk mode of operation as well. Therefore the design condition for $L_2$ remains the same in both the half cycles.

Substituting  $I_{L1v}$ in (\ref{eqn:17}),
\begin{align}
	I_{dc}&=\frac{DT_sV_{dc}}{2}\Big[\frac{D}{L_2}-\frac{1-D-D_0}{L_1}\Big]+\frac{DT_sV_{dc}}{2L_1}(1-D_0) \nonumber \\
	&=\frac{D^2T_sV_{dc}}{2L_{eq}}. \label{eqn:24}
\end{align}
Since the output voltage and current is sinusoidal and they are in phase with each other,
\begin{equation}
	I_{dc}(t)=2I_{pv}\sin^2\omega t \label{eqn:25}
\end{equation} wherein, $I_{pv}$ is the average current drawn from PV module. Further,  $V_{pv}=V_{dc}$ and $D(t)$ being positive, its expression can be written as 
\begin{align}
	\begin{gathered}
	 D(t)=D_{peak}|\sin\omega t|\quad
		\text{where,}\quad D_{peak}=2\sqrt{\frac{I_{pv}L_{eq}}{T_sV_{pv}}}
	\end{gathered}\label{eqn:d}
\end{align}
Hence $D(t)$ needs to be a rectified sine wave with an amplitude of $D_{peak}$. The value of $D_{peak}$ will be dictated by the irradiance level experienced by the solar PV module.
\section{Control Configuration}
\begin{figure}[!h]
	\includegraphics[width=\linewidth]{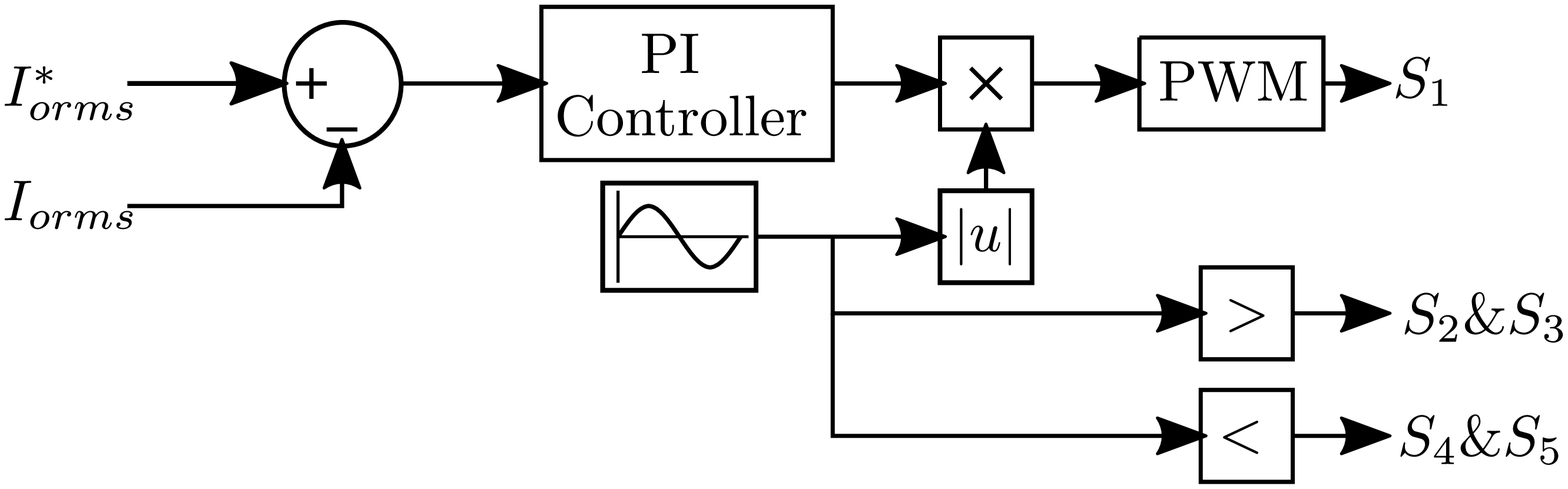}
	\caption{Block diagram of the control configuration}
	\label{fig:CC}
\end{figure}

In the Fig. \ref{fig:CC} the schematic of the control block diagram is shown. The rms value of the output current is measured and compared with the reference, $I_{orms}^*$, which is to be generated by the MPPT controller. The PI controller processes the error so obtained to determine $D_{peak}$. A rectified sinusoidal signal having an amplitude of unity is multiplied with $D_{peak}$ to obtain the instantaneous duty ratio, $D(t)$. The reference sine wave is generated by employing a single phase PLL. Switching pulses for $S_1$ are obtained by comparing $D(t)$ with a triangular carrier wave.

Switches $S_2$ and $S_3$ are turned on during the positive half cycle of the reference sine wave to operate the circuit in SEPIC mode, while $S_4$ and $S_5$ are turned on during its negative half cycle to operate the circuit in \'Cuk mode.

\section{Selection of Parameters}
\subsection{Design of input inductor ($L_1$)}
At the limiting condition, when the converter operates at the verge of discontinuous and continuous mode of operation at the peak of the grid voltage and operating at the rated condition,
\begin{align}
	D\leq\frac{v_o}{v_o+V_{dc}}=\frac{V_{om}}{V_{om}+V_{pv}}
\end{align} where, $V_{om}$ is the amplitude of the grid voltage, $v_o$. At this instant, the switching time average of $i_{L1}$ is $2I_{pv}$. Hence,
\begin{align}
	2I_{pv}&=\dfrac{I_{L1p}}{2}=\dfrac{DT_sV_{pv}}{2L_1}\nonumber\\
	\Rightarrow L_1&=\dfrac{DT_sV_{pv}}{4I_{pv}}\leq\dfrac{T_sV_{om}V_{pv}}{4I_{pv}(V_{pv}+V_{om})}
\end{align}

\subsection{Design of intermediate capacitor ($C_1$)}
As the average voltage of $v_{C1}$ is less in SEPIC mode of operation, the voltage ripple of $C_1$ is much higher in this mode compared to the operation of the inverter in the \'Cuk mode. Hence the design of $C_1$ is carried out considering the operation of the converter in SEPIC mode which is as follows,
\begin{align}
	C_1&=\dfrac{\Delta Q_{C1}}{\Delta V_{C1}}=\dfrac{(1-D-D_0)T_sI_{L1p}}{2V_{pv}(\frac{\Delta V_{C1}}{V_{C1}})}=\frac{(DT_s)^2V_{pv}}{2V_oL_1(\frac{\Delta V_{C1}}{V_{C1}})}.
\end{align}
\subsection{Design of output inductor ($L_2$)}
The switching time average current through $L_2$ is the current fed to the grid during this interval. The current ripple in $L_2$ is derived and subsequently the design criterion of $L_2$ is obtained as follows,
\begin{align}
	\Delta I_{L2}=\dfrac{DT_sV_{dc}}{L_2}
	\Rightarrow L_2=\dfrac{DV_{pv}T_s}{I_o(\frac{\Delta I_{L2}}{I_{L2}})}.
\end{align}
\subsection{Design of output capacitor ($C_2$)}
The output capacitor is chosen so that the cut off frequency, $f_c$ of the $L_2-C_2$ filter is way above the grid frequency, $f_g$, and way below the switching frequency, $f_s$, i.e. $f_g<<f_c<<f_s$ and the value of $C_2$ is obtained to be,
\begin{equation}
	C_2=\frac{1}{4\pi^2f_c^2L_2}
\end{equation}

\section{Simulated Performance of the Proposed Converter}
\begin{table}[!h]
	\centering
	\caption{Simulation Parameters: (a)~system parameters and (b)~converter parameters}
	\label{tab:table1}
	\begin{minipage}{0.48\linewidth}
		%		\centering
		\begin{tabular}{llr}
			\hline
			\\[-0.8em]
			\textbf{Parameters}  &  & \textbf{Values}    \\ \hline \hline
			\\[-0.8em]
			$V_{dc}$ &  & 35 V \\
			$V_o$    &  & 220 V    \\
			$f_g$    &  & 50 Hz    \\
			$f_s$    &  & 100 kHz  \\
			$R_o$	 & & 194 $\Omega$\\ \hline
		\end{tabular}
		\subcaption{}
	\end{minipage}
	\hfil
	\begin{minipage}{0.48\linewidth}
		\centering
		\begin{tabular}{llr}
			\hline
			\\[-0.8em]
			\textbf{Parameters}   &  & \textbf{Values}                \\ \hline \hline \\[-0.8em]
			$L_1$,  $r_{L1}$&  & 8 $\mu$H, 20~m$\Omega$\\
			$L_2$,  $r_{L2}$ &  & 100 $\mu$H, 0.6~$\Omega$\\
			$C_1$,  $r_{C1}$ &  & 0.47 $\mu$F, 30~m$\Omega$\\
			$C_2$,  $r_{C2}$ &  & 0.47 $\mu$F, 30~m$\Omega$\\
			$L_g$ &  & 1 mH\\ \hline
		\end{tabular}
		\subcaption{}
	\end{minipage}
	%	\begin{tabular}{|l r|l r|}
	%		\hline
	%		\textbf{System Parameters} & & \textbf{Converter Parameters} & \\
	%		\hline
	%		$V_{dc}$&35~V & $L_1$ & 8~$\mu$H \\
	%		$V_{o}$&220~V & $C_1$ & 1~$\mu$F\\
	%		$f_{g}$&50~Hz & $L_2$ & 100~$\mu$H\\
	%		$R_o$&50~$\Omega$ & $C_2$ & 3~$\mu$F\\
	%		$f_{sw}$&100~kHz & $L_f$ & 100~$\mu$H\\
	%		$I_{orms}^*$&2.2~A & $C_f$ & 1~mF\\
	%		\hline
	%	\end{tabular}
\end{table}
Detailed simulation studies of the proposed converter have been carried out  on MATLAB/Simulink platform. For simplicity, the grid is replaced by a load resistance while maintaining the voltage across this resistance to be equal to 220~V ac. In a realistic grid connected system, reference output current, $I_{rms}^*$ would be determined by the MPPT controller. However, in this simplistic system $I_{rms}^*$ is obtained by employing a PI controller which maintains the terminal voltage of the inverter at 220~V. The parameters chosen for the simulation model along with the parasitic series resistances of the converter are shown in Table~\ref{tab:table1}.

%\subsection{System Parameters:}
%$V_{in}$=40~V, $V_{g}$=110~V, $f_{g}$=50~Hz, $R_o$=50~$\Omega$, $f_{sw}$ =100~kHz, $I_{orms}^*$=2.2~A.
%\subsection{Converter Parameters:} 
%\begin{table}[h!]
%	\begin{center}
%		\caption{Converter Parameters}
%		\label{tab:table2}
%		\begin{tabular}{|c|c|c|c|c|c|} 
%			\hline
%			$L_1$ ($\mu$H) & $C_1$ ($\mu$F) & $L_2$ ($\mu$H) & $C_2$ ($\mu$F) & $L_f$ ($\mu$H) & $C_f$ (mF)\\
%			\hline
%			8 & 1 & 100 & 3 & 100 & 1\\
%			\hline
%		\end{tabular}
%	\end{center}
%\end{table}

The proportional gain $K_p$ and integral gain $K_i$ values of the PI controller used for controlling $I_{rms}$ are chosen to be 0.5 and 60~s$^{-1}$ respectively.
\begin{figure}[!t]
	\includegraphics[width=0.99\linewidth]{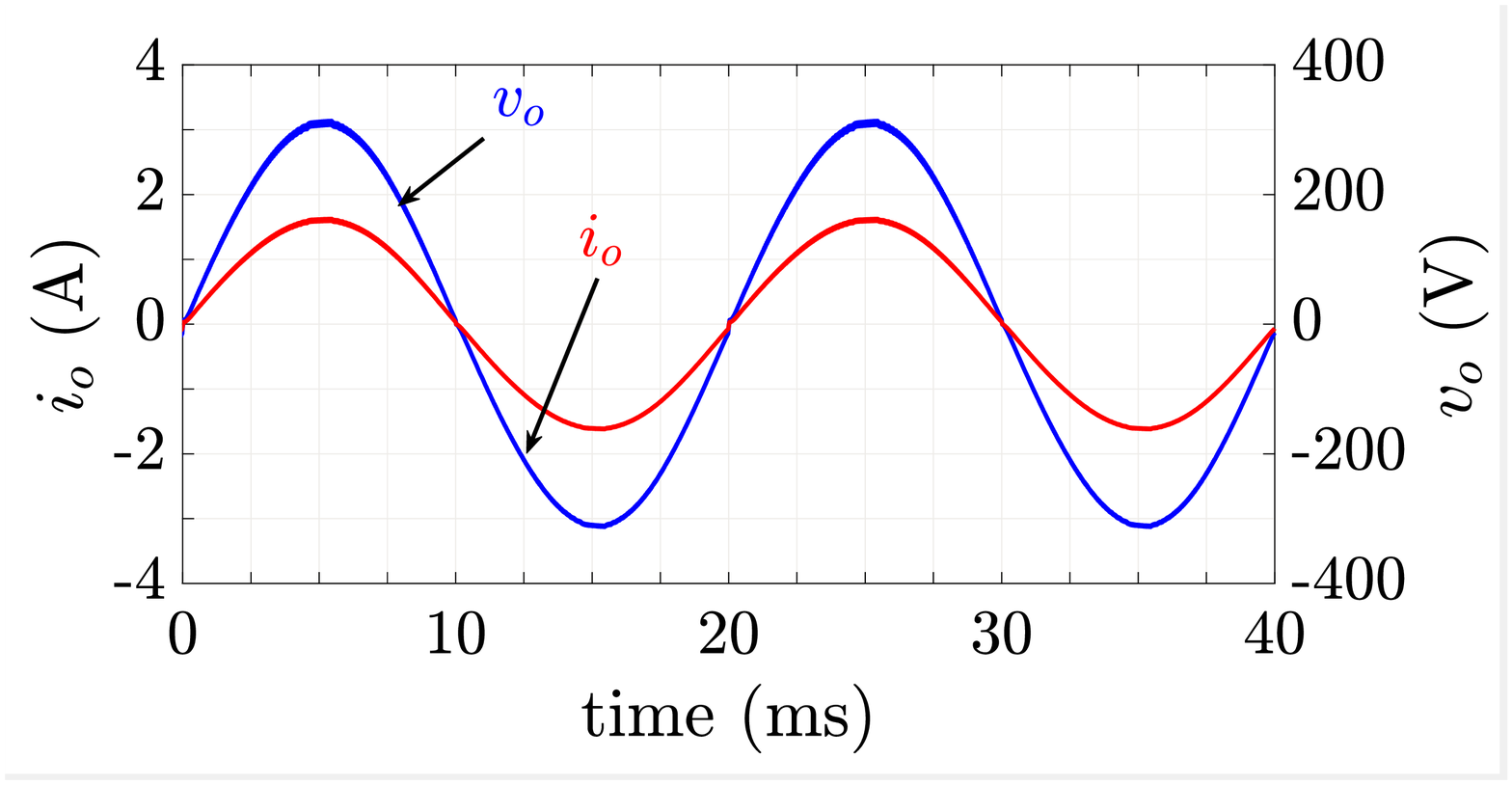}
	\caption{Output voltage ($v_o$) and current ($i_o$) waveform}
	\label{fig:vg}
\end{figure}
\begin{figure}[!t]
	\includegraphics[width=0.99\linewidth]{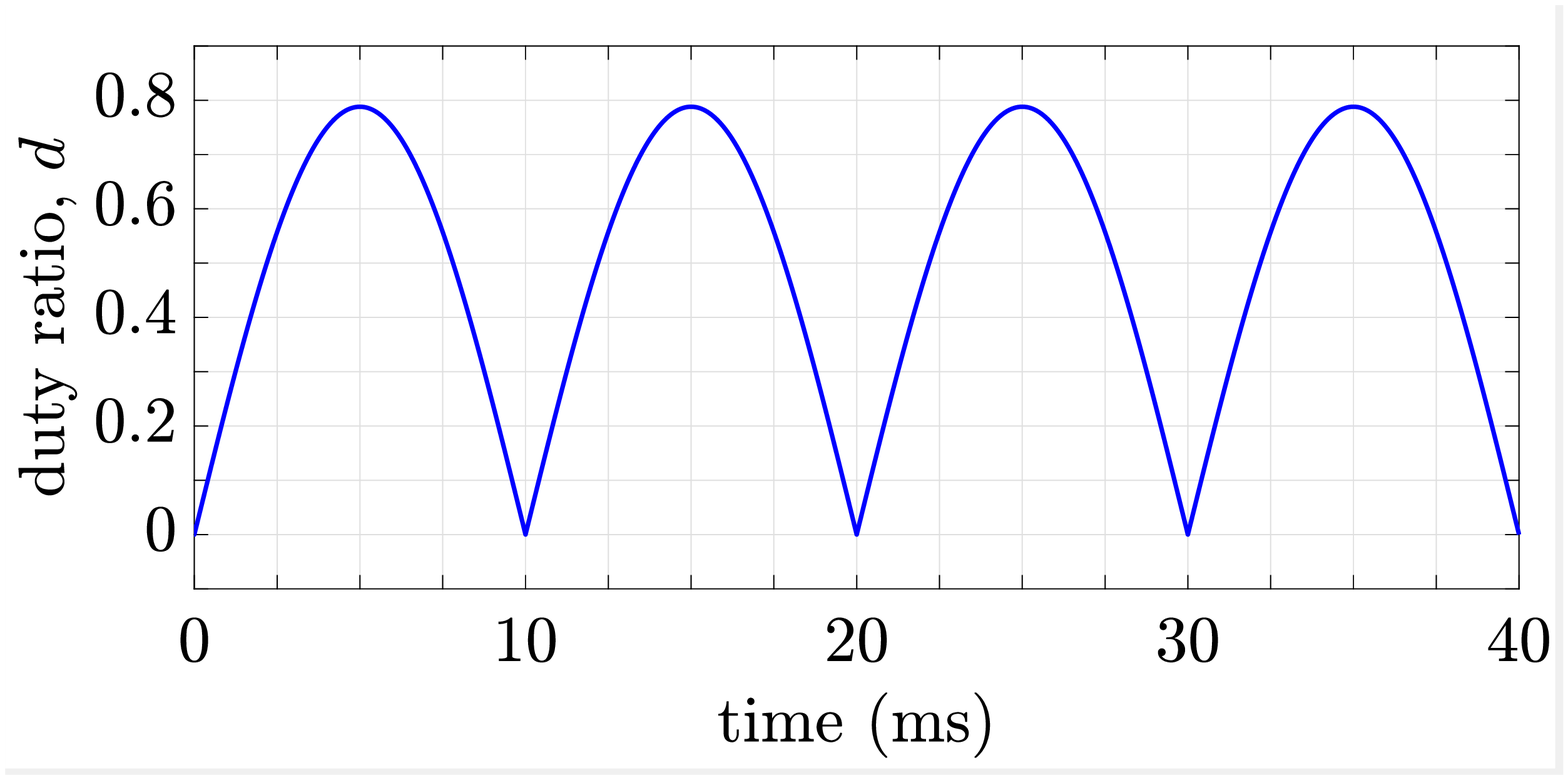}
	\caption{Waveform of the duty ratio, $d$}
	\label{fig:d}
\end{figure}
\begin{figure}[!t]
	\includegraphics[width=0.99\linewidth]{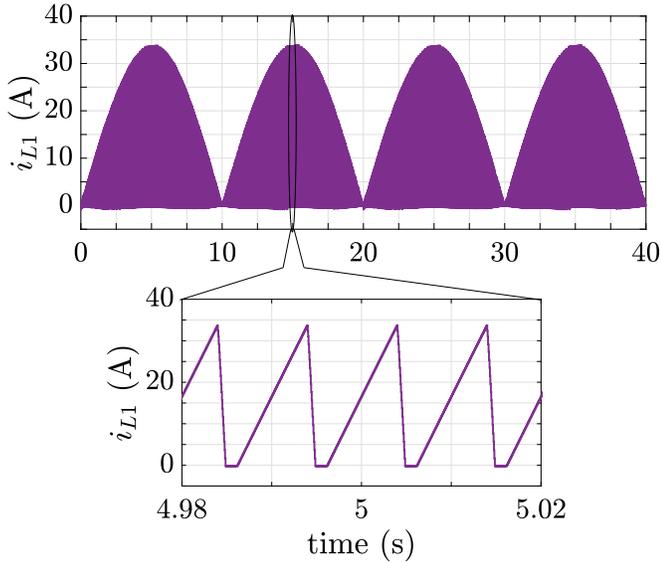}
	\caption{ Input inductor current ($i_{L1}$) over two full cycles (0.04~s) and expanded view of $i_{L1}$ from 4.98~ms to 5.02~ms}
	\label{fig:IL1}
\end{figure}
The inverter is operated with the system parameters as depicted in Table~\ref{tab:table1}(a). The steady state response of the output voltage and current of the proposed inverter is shown in Fig.~\ref{fig:vg}. It may be noted that the amplitude of the sinusoidal output voltage $v_{op}$ of the inverter is 311~V. From Fig.~\ref{fig:d} it can be inferred that the converter is operated with the peak duty ratio ($D_{peak}$) of 0.8. The THD of the output voltage is found to be 1.21\%. %The output current waveform is not shown as the load, chosen for simulation, is purely resistive.

%\begin{figure}[!h]
%	\includegraphics[width=0.99\linewidth]{./Images/IL1_1.eps}
%	\caption{(a) Input inductor current ($i_{L1}$) over one full cycle (0.02~s), (b) Expanded view of $i_{L1}$ from 4.98~ms to 5.02~ms when the $R_o$ is 250~$\Omega$}
%	\label{fig:IL1_1}
%\end{figure}
The steady state response of the input inductor current is shown in Fig.~\ref{fig:IL1}. It can be inferred from the aforementioned figures that the converter is operating in DCM.
%\begin{figure}[!h]
%	\includegraphics[width=0.99\linewidth]{./Images/Idc.eps}
%	\caption{DC side current waveform after filtering}
%	\label{fig:idc}
%\end{figure}

%The steady state current drawn from the dc source is shown in Fig. \ref{fig:idc}. It can be noted that, the current drawn from the dc source is having a dominant 2nd order harmonic (100~Hz) component, and the average current drawn from the dc side is 6.4 A.

%Though the current drawn from the DC source when the current flowing through the ind in Fig. \ref{fig:idc} is around 13 A, the maximum value of $i_{L1}$ is above 30 A for the converter is operating under DCM. This is one of the reasons behind the increment in the overall core and copper losses in the inductor $L_1$. During the designing of that inductor core, this phenomenon has to be kept in mind.

\begin{figure}[!t]
	\includegraphics[width=0.99\linewidth]{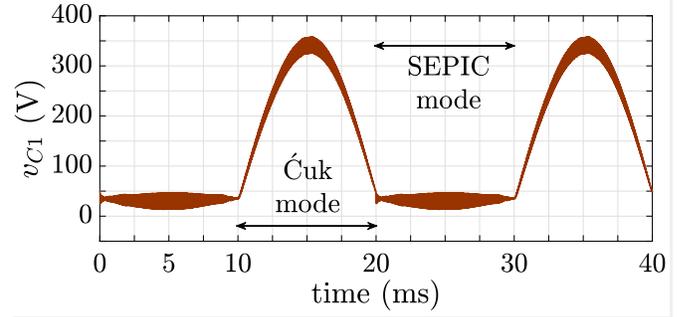}
	\caption{Intermediate capacitor voltage ($v_{C1}$) waveform}
	\label{fig:Vc1}
\end{figure}

At steady state, the voltage across the intermediate capacitor, $C_1$  is shown in Fig. \ref{fig:Vc1}. Form this figure it can be noted that the waveform of $v_{C1}$ in the positive half cycle is different from that in the negative half cycle. This difference arises as the converter operates in SEPIC mode during positive half cycle, and in \'Cuk mode during negative half cycle. %The result justifies \eqref{eqn:7} and \eqref{eqn:14}.

\begin{figure}[!t]
	\includegraphics[width=0.99\linewidth]{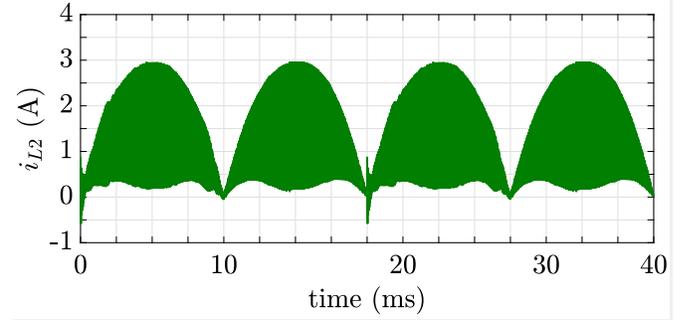}
	\caption{Output inductor current ($i_{L2}$) waveform}
	\label{fig:il2}
\end{figure}

The steady state response of output inductor current, $i_{L2}$ is shown in Fig. \ref{fig:il2}. It can be inferred that the waveform is a rectified sine wave having switching frequency harmonic components.

\begin{figure}[!t]
	\centering
	\includegraphics[width=\linewidth]{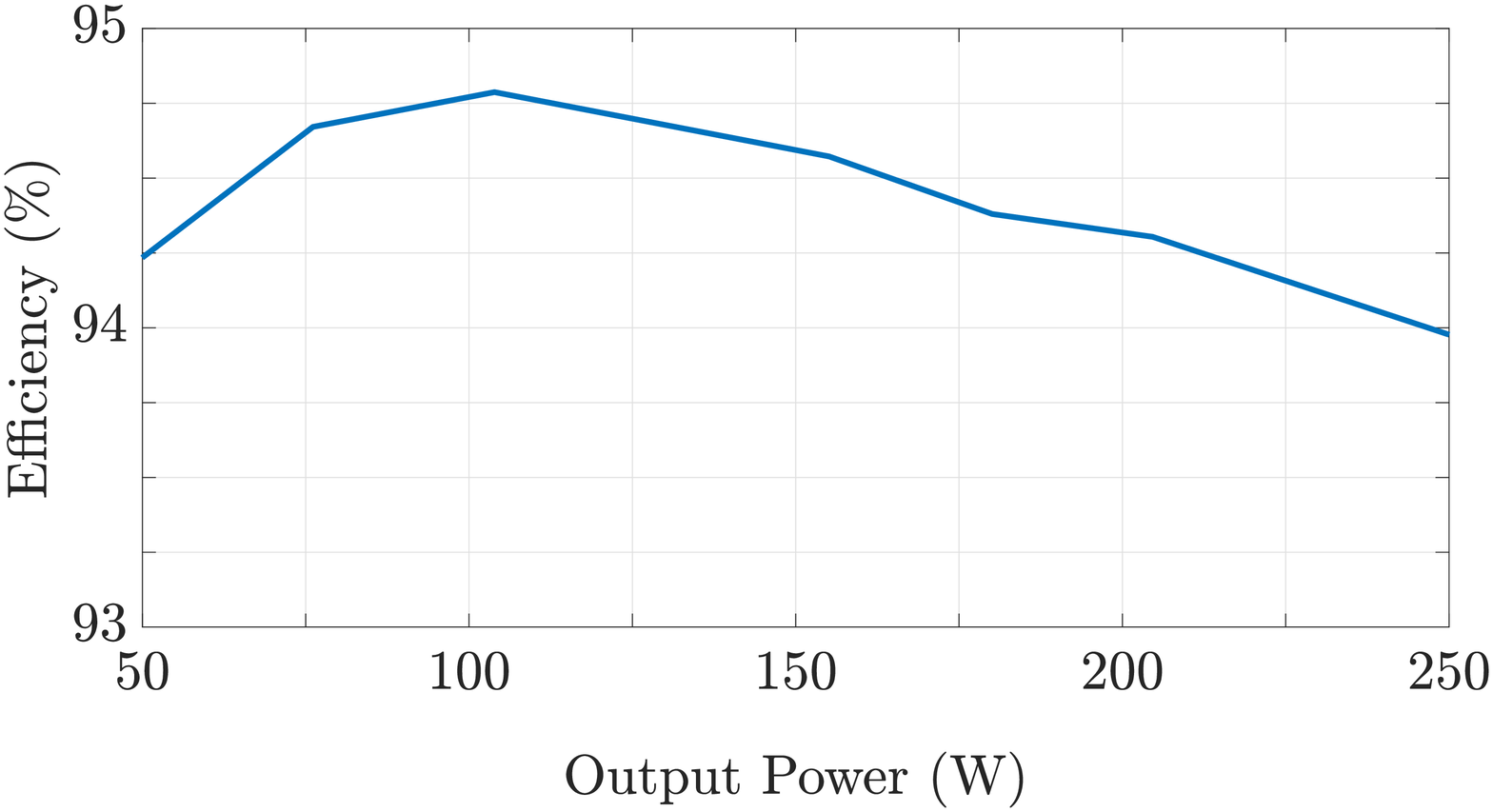}
	\caption{Estimated efficiency plot of the proposed inverter}
	\label{fig:eff}
\end{figure}

The plot of the estimated efficiency of the inverter with respect to the power level is shown in Fig.~\ref{fig:eff}. For efficiency estimation, the MOSFET, IPW60R024P7 is considered for $S_1$ and MOSFET, IPW60R037P7 is considered for $S_2-S_5$. Although the inverter is developing 220~V ac from an input voltage of 35~V dc, the efficiency of the proposed inverter is comparable with the inverters which are reported in the literature for developing 110~V ac from 35-50~V dc\cite{gautam2017design,jamatia2018cuk,rajeev2018analysis}.

%The reason behind the component having double of the line frequency in $i_{dc}$ (Fig. \ref{fig:idc}) is the instantaneous power miss-match between the ac side and dc side. The ac side power contains a constant part as well as double of the line frequency component. Both this components have to be supplied by the source, which is stiff dc in this case. So, the current drawn from the dc source contains 100~Hz component. It can be shown that, to reduce this component to 3\% of the dc component, a capacitor of 15~mF is required if the power rating is 250~W if no other special care for power decoupling is taken.

%The other method that can be used to reduce the double frequency component is available in literature~\cite{hu2013review,kyritsis2007novel}. Mostly active power decoupling strategies, in which the size of required inductor and capacitors are sufficiently small. The only compromise in this case is the use of extra HF switches, which in turn, increases the power loss. So, a proper trade of between the loss and the size of passive element is necessary.

%In this paper, the main focus was the proposed converter. So, no extra care is taken regarding power decoupling.

\section{Experimental Validation}
\begin{figure}[!t]
	\centering
	\includegraphics[width=\linewidth]{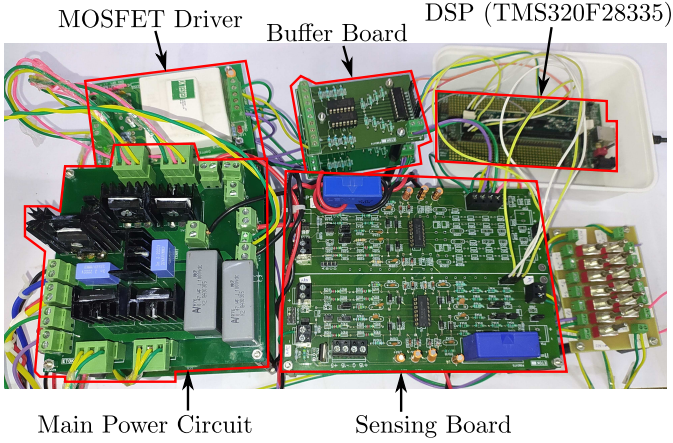}
	\caption{Photograph of the experimental prototype}
	\label{fig:hwsu}
\end{figure}

In order to confirm the the viability of the proposed inverter, a 250~W semi-engineered laboratory prototype of the micro-inverter is fabricated, and detailed experimental studies have been carried out. The passive components as mentioned in Table~\ref{tab:table1}(b) are used to realize the prototype. In order to increase the reliability, thin film capacitors are used. Switch $S_1$ is realized by the MOSFET, IPW60R024P7 while $S_2-S_5$ are realized by MOSFET, IPW60R037P7.
The controller of the inverter is realized by utilising DSP, TMS320F28355. A 1.5~kW programmable power supply (EPS power supply PSI 9360-15) is used as the input dc source. The photograph of the prototype is shown in Fig.~\ref{fig:hwsu}.

\begin{figure}[!t]
	\centering
	\includegraphics[width=\linewidth]{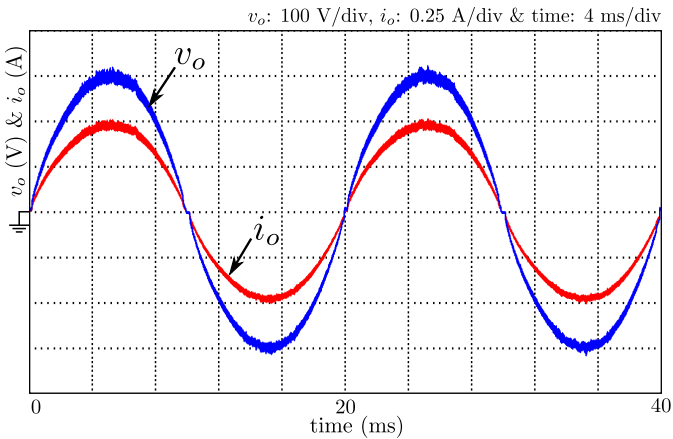}
	\caption{Experimental performance: waveform of output voltage $v_o$ and output current $i_o$}
	\label{fig:hwvo}
\end{figure}
\begin{figure}[!t]
	\centering
	\includegraphics[width=\linewidth]{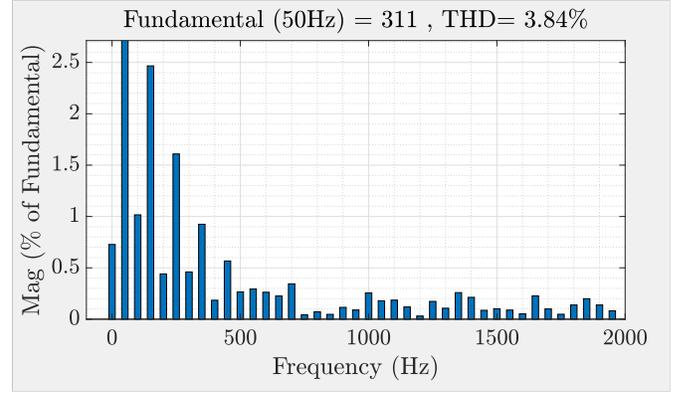}
	\caption{Experimental performance: FFT of output voltage waveform}
	\label{fig:hwTHD}
\end{figure}
\begin{figure}[!t]
\centering
\includegraphics[width=\linewidth]{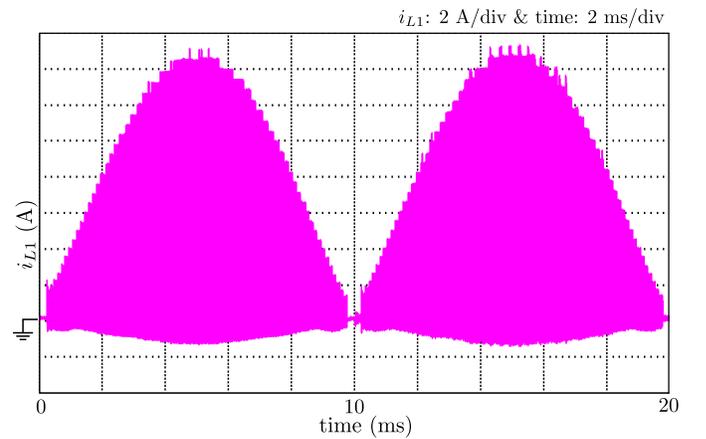}
\caption{Experimental performance: waveform of input inductor current $i_{L1}$}
\label{fig:hwil1}
\end{figure}

The measured performance of the inverter is shown Fig.~\ref{fig:hwvo} wherein the input voltage is maintained at 35~V dc and the load resistance of 660~$\Omega$ is connected across its output terminals. It can be noted that the peak of the output voltage developed across the load resistance is 311~V rms. In Fig.~\ref{fig:hwil1} the measured waveform of $i_{L1}$ is depicted which indicates that at light load condition the circuit operates in DCM. The operating efficiency of the inverter while it is made to deliver 73~W is found to be 89.5\%. The THD in the output voltage is 3.84\% (Fig.~\ref{fig:hwTHD}) which is well within the stipulated standards.

\begin{figure}[!t]
	\centering
	\includegraphics[width=\linewidth]{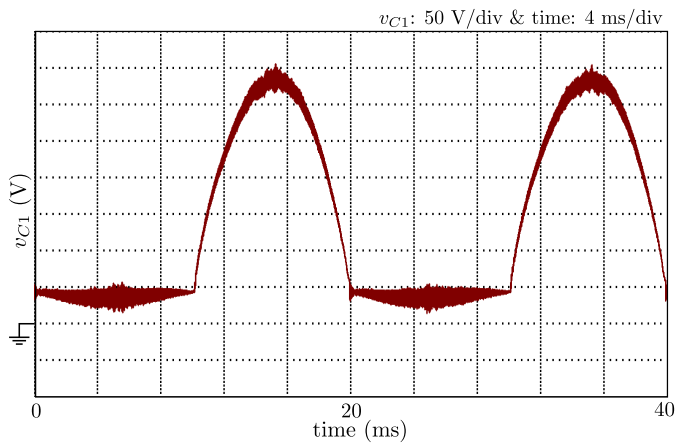}
	\caption{Experimental performance: voltage waveform across intermediate capacitor $C_1$}
	\label{fig:hwvc}
\end{figure}
The measured voltage across $C_1$ while the inverter is operating at steady state is shown in Fig.~\ref{fig:hwvc} which resembles the simulated waveform of Fig.~\ref{fig:Vc1}. Since in the positive half cycle, $V_{C1}=V_{dc}$, it can be inferred from Fig.~\ref{fig:hwvc} that the input voltage is maintained at 35~V. In negative half cycle the peak value of the waveform is around 350~V since it is the sum of the input voltage and peak of the sinusoidal output voltage, $v_o$.

\begin{figure}[!t]
	\centering
	\includegraphics[width=\linewidth]{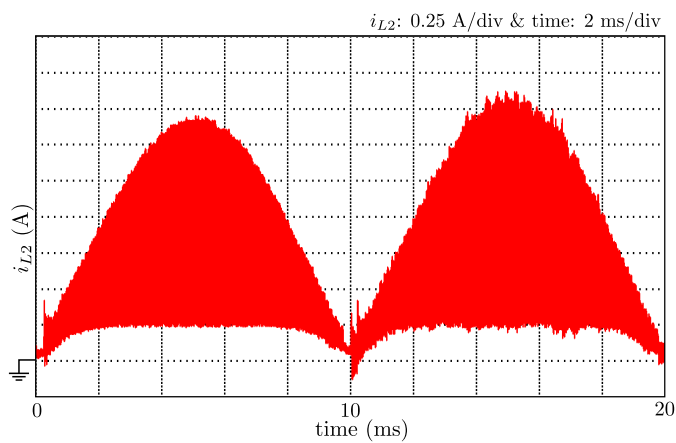}
	\caption{Experimental performance: waveform of output inductor current $i_{L2}$}
	\label{fig:hwil2}
\end{figure}
The measured steady state current through $L_2$ is shown in Fig~\ref{fig:hwil2}. It may be noted that it is polluted with high frequency switching harmonics. It also shows that for this operating condition $I_{L2v}$ remains to be positive. However after getting manipulated by the the unfolding and filtering circuits the output current, $i_o$ becomes almost sinusoidal current (Fig.~\ref{fig:hwvo}) having a THD of 3.84\%. 

\section{Conclusion}
A combined SEPIC-\'Cuk based micro-inverter topology has been proposed in this paper. The salient features of the proposed inverter are as follows,
\begin{itemize}
	\item it operates in SEPIC mode for positive half cycle and in \'Cuk mode for negative half cycle
	\item the inverter is realized by using single high frequency switch thereby improving its reliability and cost
	\item four line frequency switches are employed to interchange the modes between SEPIC and \'Cuk mode, and the switching losses associated with these switches are negligible
	\item in order to obtain high gain the inverter is made to operated in DCM. The DCM operation also ensures that the turn on loss of the high frequency switch is negligible
	\item the neutral of the PV module is shorted with that of the grid thereby eliminating the flow of leakage current
\end{itemize}

The operating principle, detailed mathematical modelling of the system, design guidelines for the passive elements, control strategy are presented in the paper. The viability of the micro-inverter is validated by carrying out detailed simulation and experimental studies.

%\appendices
%\section{Proof of the First Zonklar Equation}
%Appendix one text goes here.
%\section{}
%Appendix two text goes here.
%
%\section*{Acknowledgment}
%The authors would like to thank...

\bibliographystyle{IEEEtran}
\bibliography{IEEEabrv,ref}
% <OR> manually copy in the resultant .bbl file
% set second argument of \begin to the number of references
% (used to reserve space for the reference number labels box)
%\begin{thebibliography}{1}
%
%\bibitem{IEEEhowto:kopka}
%H.~Kopka and P.~W. Daly, \emph{A Guide to \LaTeX}, 3rd~ed.\hskip 1em plus
%  0.5em minus 0.4em\relax Harlow, England: Addison-Wesley, 1999.
%
%\end{thebibliography}

% biography section

%\begin{IEEEbiography}[{\includegraphics[width=1in,height=1.25in,clip,keepaspectratio]{mshell}}]{Michael Shell}

%\begin{IEEEbiography}{Michael Shell}
%Biography text here.
%\end{IEEEbiography}
%
%\begin{IEEEbiographynophoto}{John Doe}
%Biography text here.
%\end{IEEEbiographynophoto}
%
%\begin{IEEEbiographynophoto}{Jane Doe}
%Biography text here.
%\end{IEEEbiographynophoto}

\end{document}